\documentclass[prl,twocolumn,superscriptaddress,showpacs]{revtex4}
\usepackage{graphicx,amsmath,amssymb,bm,multirow}

\newcommand{\be}{\begin{equation}}
\newcommand{\ee}{\end{equation}}
\newcommand{\bea}{\begin{eqnarray}}
\newcommand{\eea}{\end{eqnarray}}
\newcommand{\ba}{\begin{align}}
\newcommand{\ea}{\end{align}}

\newcommand{\fmi}{\, \text{fm}^{-1}}

\newcommand{\gcmiq}{\, \text{g} \, \text{cm}^{-3}}
\newcommand{\mev}{\, \text{MeV}}
\newcommand{\vlowk}{V_{{\rm low}\,k}}

\newcommand{\om}{\omega}
\newcommand{\qq}{{\bf q}}
\newcommand{\kf}{k_{\rm F}}
\newcommand{\pp}{{\bf p}}
\newcommand{\kk}{{\bf k}}

\newcommand{\wxchi}{\widetilde{X}_\sigma}

\begin{document}

\title{Chiral effective field theory calculations of neutrino processes 
in dense matter}

\author{S.\ Bacca}
\email[E-mail:~]{bacca@triumf.ca}
\affiliation{TRIUMF, 4004 Wesbrook Mall, Vancouver, BC, V6T 2A3, Canada}
\author{K.\ Hally}
\email[E-mail:~]{079203h@acadiau.ca}
\affiliation{TRIUMF, 4004 Wesbrook Mall, Vancouver, BC, V6T 2A3, Canada}
\affiliation{Acadia University, Department of Physics, P.O.\ Box 49,
Wolfville, Nova Scotia, B4P 2R6, Canada}
\author{C.\ J.\ Pethick}
\email[E-mail:~]{pethick@nbi.dk}
\affiliation{The Niels Bohr Institute, Blegdamsvej 17, DK-2100
Copenhagen \O, Denmark}
\affiliation{NORDITA, Roslagstullsbacken 21, 10691 Stockholm, Sweden}
\author{A.~Schwenk}
\email[E-mail:~]{schwenk@triumf.ca}
\affiliation{TRIUMF, 4004 Wesbrook Mall, Vancouver, BC, V6T 2A3, Canada}


\begin{abstract}
We calculate neutrino processes involving two nucleons
at subnuclear densities using chiral effective field theory.
Shorter-range noncentral forces reduce the neutrino rates significantly
compared with the one-pion exchange approximation currently used in 
supernova simulations. For densities $\rho < 10^{14} \gcmiq$, we find
that neutrino rates are well constrained by nuclear interactions and
nucleon-nucleon scattering data. As an application, we calculate the 
mean-square energy transfer in scattering of a neutrino from nucleons
and find that collision processes and spin-dependent mean-field effects
dominate over the energy transfer due to nucleon recoil.
\end{abstract}

\pacs{97.60.Bw, 26.50.+x, 95.30.Cq, 26.60.-c}

\maketitle

Neutrino processes involving two nucleons (NN) play a special role
in the physics of core-collapse supernovae and neutron stars: 
Neutrino-pair bremsstrahlung and absorption, $N N 
\leftrightarrow N N \nu \overline{\nu}$, are key for the production
of muon and tau neutrinos, for their spectra, and for equilibrating 
neutrino number densities~\cite{Raffelt,Suzuki,Raffelt1,HR,TBH,KRJ}.
Since neutrinos interact weakly, the rates for neutrino
emission, absorption and scattering are determined by the dynamic
response functions of strongly-interacting matter. Supernova 
explosions are most sensitive to neutrino processes near the 
protoneutron star, at subnuclear densities $\rho \lesssim \rho_0 /10$ 
($\rho_0 = 2.8 \times 10^{14} \gcmiq$ being the saturation density),
and where matter is neutron rich. In this regime, there exist
to date no systematic calculations of neutrino rates that go 
beyond the one-pion exchange (OPE) approximation for the 
nucleon-nucleon interaction. In this paper, we present first
results for neutrino processes in neutron matter based on chiral 
effective field theory (EFT).

Noncentral contributions to strong interactions, due to tensor 
forces from pion exchanges and spin-orbit forces, are essential
for the two-nucleon response. This follows from direct calculations of
neutrino-pair bremsstrahlung~\cite{FM} and from conservation 
laws~\cite{OP}. In supernova and neutron star simulations, the 
standard rates for bremsstrahlung and absorption are based on
the OPE approximation~\cite{FM,HR}.
This is a reasonable starting point, since it represents the 
long-range part of nuclear forces, and for neutron matter, it 
is the leading-order contribution in chiral EFT~\cite{Epelbaum}.
However, for the relevant Fermi momenta $\kf \sim 1.0 \fmi \approx
200 \mev$, subleading noncentral contributions are crucial for
reproducing NN scattering data~\cite{Epelbaum}. In this paper, we 
go beyond the OPE approximation and include 
contributions up to next-to-next-to-next-to-leading order 
(N$^3$LO) in chiral EFT. We find that shorter-range noncentral
forces significantly reduce the neutrino rates for all relevant
densities. As an application, we calculate the mean-square energy
transfer in scattering of a neutrino from nucleons and find that
collision processes and spin-dependent mean-field effects are
more important than nonzero momentum transfers to the nucleons.
This establishes that the long-wavelength approximation
used in the OPE calculations~\cite{HR} is reasonable.

We follow the approach to neutrino processes in nucleon matter
developed in Ref.~\cite{LPS}, which is based on Landau's theory
of Fermi liquids and consistently includes one-quasiparticle-quasihole
pair states (corresponding to elastic scattering of neutrinos from
nucleons) and two-quasiparticle-quasihole pair states, which are 
taken into account through the collision integral in the Landau 
transport equation for quasiparticles. Using a 
relaxation time approximation, the transport equation can be
solved and this leads to a general form for the response 
functions~\cite{LPS}. The spin response includes 
multiple-scattering effects, thereby taking into account the 
Landau-Pomeranchuk-Migdal (LPM) effect, and generalizes earlier work
to finite wavelengths.

\begin{figure}[b]
\begin{center}
\includegraphics[scale=0.34,clip=]{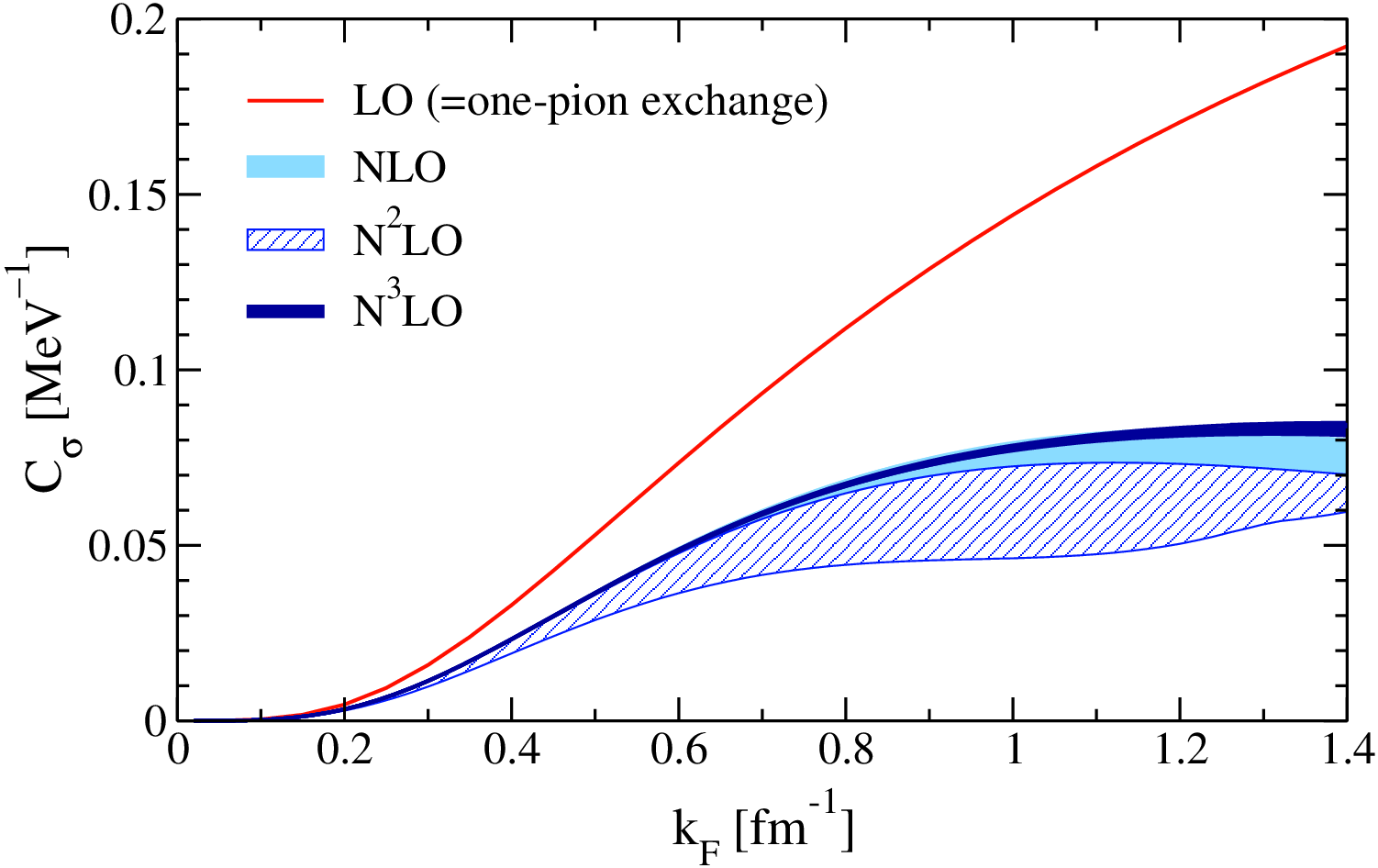}
\end{center}
\vspace*{-3mm}
\caption{(Color online) Spin relaxation rate given by $C_\sigma$
of Eq.~(\ref{strace}) as a function of Fermi momentum $\kf$ obtained
from chiral EFT interactions of successively higher orders~\cite{EGM}.
All results are for $m^*/m = 1$.
\label{orders}}
\end{figure}

{\it Basic formalism.}
In supernovae and neutron stars, the energy $\om$ and momenta 
$\qq$ transferred by neutrinos to the system are small compared
to the momenta of nucleons. In addition, we consider degenerate
conditions, where the temperature is small compared to the 
Fermi energy, $T/\varepsilon_{\rm F} \lesssim 1/3$. This is 
the regime in which Landau's theory of Fermi liquids is a 
reasonable first approximation, and includes the conditions
under which the two-nucleon response is effective. We focus
on axial current processes, since they dominate in many 
situations and are most affected by NN interactions. The
corresponding neutrino rates are given by~\cite{Raffelt}
\be
\Gamma(\om,\qq)
= 2\pi \, n \, G_{\rm F}^2 \, C_{\rm A}^2 \, (3-\cos\theta) \, 
S_{\rm A}(\om,\qq) \,,
\ee
multiplied by occupation probabilities for the initial neutrino
states and Pauli-blocking factors for final states. Here $n$ 
denotes the neutron number density, $G_{\rm F}$ the Fermi
coupling constant, $C_{\rm A} = - g_a/2 = - 1.26/2$ the axial-vector
coupling for neutrons, and $\theta$ is the angle between the initial
and final neutrino momenta for scattering, or between the neutrino
momentum and minus the antineutrino momentum for neutrino-pair
bremsstrahlung and absorption. The axial or spin dynamical 
structure factor $S_{\rm A}$ is given by~\cite{IP}
\be
S_{\rm A}(\om,\qq) = \frac{1}{\pi n} \, \frac{1}{1-e^{-\om/T}} \,
{\rm Im} \, \chi_\sigma(\om,{\bf q}) \,,
\label{structspin}
\ee
where $\chi_\sigma$ is the spin-density--spin-density response 
function, and we use units with $\hbar=c=k_{\rm B} = 1$. The 
solution~\cite{LPS} to the Landau transport equation leads to
$ \chi_\sigma = N(0) \,  \wxchi/(1+ G_0 \wxchi)$,
with density of states at the Fermi surface $N(0)=m^* \kf / \pi^2$, 
nucleon effective mass $m^*$, and Landau parameter $G_0$ for the 
spin-dependent part of the interaction.
In the relaxation time approximation, the response function $\wxchi$
(for $G_0=0$) is
\be
\wxchi = 1 - \frac{\om}{2 v_{\rm F} q} \ln \biggl(
\frac{\omega+ i/\tau_\sigma + v_{\rm F} q}{\omega+ i/\tau_\sigma - v_{\rm F} q}
\biggr) \,, 
\label{Xsigma}
\ee
where $\tau_\sigma$ denotes the spin relaxation time and 
$v_{\rm F}=\kf / m^*$ is the Fermi velocity.
For frequencies $|\omega| \gg 1/\tau_\sigma$ and $q \ll \kf$,
the spin relaxation rate is given 
by~\cite{LPS,LOP}
\begin{widetext}
\be
\frac{1}{\tau_\sigma} = C_\sigma \, \bigl[
T^2 + (\om/2\pi)^2 \bigr] \quad \text{with} \quad
C_\sigma = \frac{\pi^3 m^*}{6 \kf^2} \: \biggl\langle \:
\frac{1}{12} \, \sum\limits_{k=1,2,3}
{\rm Tr} \biggl[ \, {\cal A}_{{\bm \sigma}_1,{\bm \sigma}_2}(\kk,\kk') \,
{\bm \sigma}_1^k \bigl[ ({\bm \sigma}_1 + {\bm \sigma}_2)^k \, , \,
{\cal A}_{{\bm \sigma}_1,{\bm \sigma}_2}(-\kk,\kk') \bigr] \,
\biggr] \biggr\rangle \,,
\label{strace}
\ee
\end{widetext}
where ${\cal A}_{{\bm \sigma}_1,{\bm \sigma}_2}(\kk,\kk')$ is the
quasiparticle scattering amplitude multiplied by $N(0)$,
$\kk = \pp_1 - \pp_3$ and $\kk' = \pp_1 - \pp_4$ are the nucleon
momentum transfers, and the average $\langle \ldots \rangle$ is over
the Fermi surface (for details see~\cite{LPS}). The commutator
with the two-body spin operator demonstrates that only noncentral
interactions contribute.

We treat the strong interaction in Born approximation and evaluate
the spin trace for $C_\sigma$ in two-body spin space $| s \, m_s
\rangle$ using a partial-wave expansion. For $s=0$ the spin trace 
vanishes, thus only $s=1$ states
and odd orbital angular momenta $\ell, \ell', \tilde{\ell}, \tilde{\ell'}$
contribute, and we find
\begin{widetext}
\begin{align}
C_\sigma &= \frac{16 \pi m^{* 3}}{9 \kf} \sum_{j \ell \ell'} \,
\sum_{\tilde{j} \tilde{\ell} \tilde{\ell'}} \, \sum_L 
\sum_{J={\rm even}} \, \sum_{m_{s}m'_{s}}
i^{\ell-\ell'+\tilde{\ell'}-\tilde{\ell}} \,
(-1)^{j+\tilde{j}+L} \, \bigl( \, \widehat{j} \, \widehat{\tilde{j}} \,
\widehat{L} \, \widehat{J} \, \bigr)^2 \, \widehat{\ell} \, \widehat{\ell'} \,
\widehat{\tilde{\ell}} \, \widehat{\tilde{\ell'}} \, \biggl[\frac{J!}{2^J \,
(J/2)!^2}\biggr]^2
\left(
\begin{array}{c c c}
\ell & \tilde{\ell} & J \\
0 & 0 & 0 \\
\end{array}
\right)
\left(
\begin{array}{c c c}
\ell' & \tilde{\ell'} & J \\
0 & 0 & 0 \\
\end{array}
\right) \nonumber \\
& \times
\left\{
\begin{array}{c c c}
\ell & \ell' & L \\
1 & 1 & j \\
\end{array}
\right\}
\left\{
\begin{array}{c c c}
\tilde{\ell'} & \tilde{\ell} & L \\
1 & 1 & \tilde{j} \\
\end{array}
\right\}
\left\{
\begin{array}{c c c}
\ell' & \ell & L \\
\tilde{\ell} & \tilde{\ell'} & J \\
\end{array}
\right\}
\biggl[ {\cal C}_{L (m_{s}-m'_{s})1m'_{s}}^{1m_{s}} \biggr]^{2}
(m_{s}^{2}-m_{s}m'_{s}) 
\int_0^{\kf} \frac{p \, dp}{\sqrt{\kf^2 - p^2}} \,
\langle p |V_{\ell' \ell}^{j s=1}| p \rangle
\langle p |V_{\tilde{\ell} \tilde{\ell}'}^{\tilde{j} s=1}| p \rangle \,,
\label{pw}
\end{align}
\end{widetext}
where $j, \tilde{j}$ are total angular momenta, $p$ is the magnitude of
the relative momenta $p = |{\bf p}_1 - {\bf p}_2|/2 = |{\bf p}_3 - 
{\bf p}_4|/2$, and $\langle p |V_{\ell' 
\ell}^{j s}| p \rangle$ are partial-wave matrix elements of the
strong interaction.
We use $\widehat{a}=\sqrt{2a+1}$ and standard notation
for Clebsch-Gordan, $3j$ and $6j$ symbols~\cite{angular}.

{\it Results.} In Fig.~\ref{orders}, we show the spin relaxation
rate calculated from chiral EFT interactions up to N$^3$LO as a function
of the Fermi momentum over a wide density range ($\kf 
= 1.4 \fmi$ corresponds to $\rho = m \kf^3/(3 \pi^2) = 1.6 \times 
10^{14} \gcmiq$)~\footnote{For all results, we found excellent
convergence with partial waves $j, \tilde{j} \leqslant 6$. In 
addition, we checked that the spin relaxation rate for OPE 
expanded in partial waves reproduces the analytical result
(Eq.~(38) in Ref.~\cite{LPS}).}.  
The neutrino-pair bremsstrahlung and absorption
rates are proportional to
$C_\sigma$ to a good approximation when $|\omega| \gg 1/\tau_\sigma$,
while for larger values of $C_\sigma$ they are suppressed by the LPM
effect~\cite{Raffelt,LPS}. 
The leading order (LO) contribution includes 
OPE as the only noncentral interaction, which provides the standard
two-nucleon rates used in current supernova simulations. Our results
based on chiral EFT interactions of successively higher orders 
of Epelbaum {\it et al.}~(EGM)~\cite{EGM} show that OPE significantly 
overestimates $C_\sigma$ for all relevant densities. The bands at
next-to-leading order (NLO), N$^2$LO, and N$^3$LO provide an
estimate of the theoretical uncertainty, generated by varying
the cutoff $\Lambda$ as well as a spectral function
cutoff in the irreducible $2\pi$-exchange $\Lambda_{\rm 
SF}$~\cite{Epelbaum,EGM}. Chiral EFT interactions at N$^3$LO
accurately reproduce low-energy NN scattering~\cite{EGM,EM}, 
and at this order, $C_\sigma$ is practically independent of the 
N$^3$LO potential for these densities.
Most of the reduction of $C_\sigma$ occurs at the NLO level,
which includes the leading $2\pi$-exchange tensor force
(for $\Lambda_{\rm SF} \to \infty$)
\be
V^{\rm NLO}_{2\pi,\,{\rm t}} = - \, {\bm \sigma}_1 \cdot {\bf k} \:
{\bm \sigma}_2 \cdot {\bf k} \: \frac{3 \, w \, g^4_a}{
64 \, k \, \pi^2 F^4_{\pi}} \, \ln \frac{w+k}{2m_{\pi}} \,,
\ee
and shorter-range noncentral contact interactions
\begin{multline}
V^{\rm NLO}_{C,\,{\rm nc}} = 
\widetilde{C}_1 \: {\bm \sigma}_1 \cdot {\bf k} \:
{\bm \sigma}_2 \cdot {\bf k}
+ \widetilde{C}_2 \: {\bm \sigma}_1 \cdot {\bf k}' \:
{\bm \sigma}_2 \cdot {\bf k}' \\
+ \widetilde{C}_3 \: i \, ({\bm \sigma}_1 - {\bm \sigma}_2)
\cdot ({\bf k} \times {\bf k}') \,,
\end{multline}
where $w=\sqrt{4m^2_{\pi}+k^2}$ and $\widetilde{C}_i$ are constrained
by NN scattering data (S- and P-waves). $V^{\rm NLO}_{2\pi,\,{\rm t}}$
added to OPE increases $C_\sigma$ (to $0.29 \mev^{-1}$ at $\kf = 1.0
\fmi$), but this is compensated by the repulsive shorter-range
$V^{\rm NLO}_{C,\,{\rm nc}}$.

\begin{figure}[t]
\begin{center}
\includegraphics[scale=0.435,clip=]{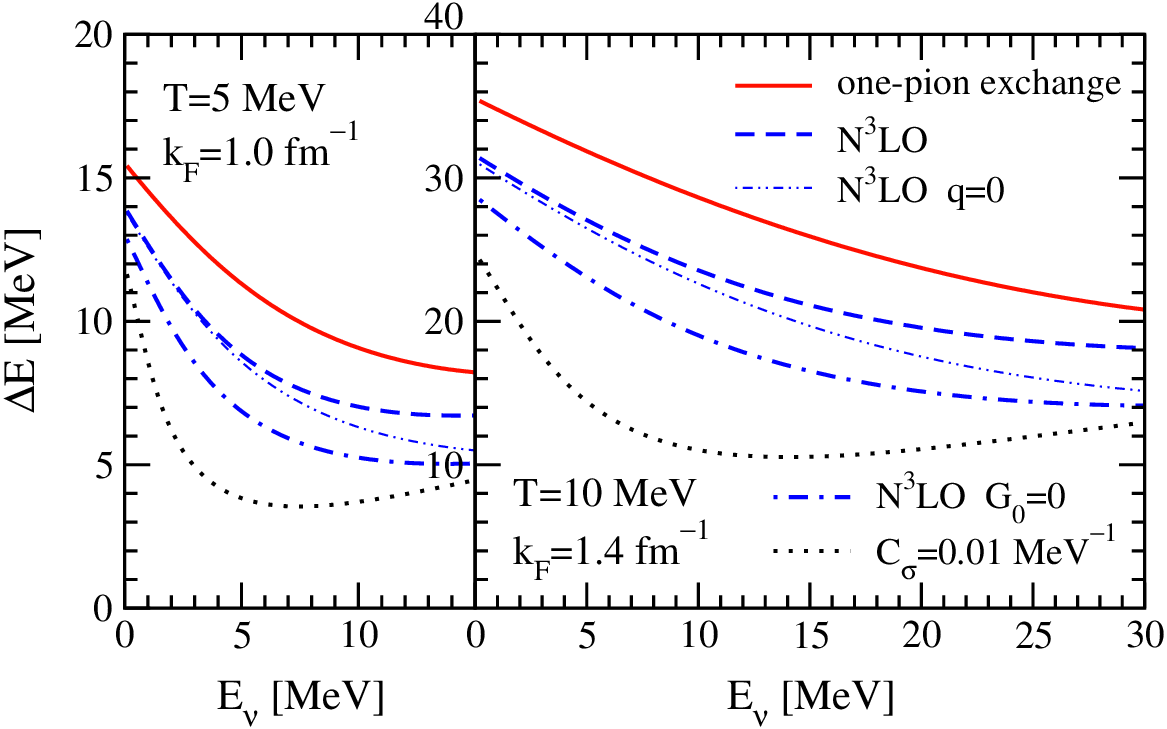}
\end{center}
\caption{(Color online) The rms energy transfer $\Delta E$ in neutrino 
scattering from neutrons as a function of initial neutrino energy $E_\nu$
for two $T$-$\kf$ combinations. Results are shown for different $C_\sigma$
(OPE, N$^3$LO using Eq.~(\ref{fit}), 
and a lower value), in the long-wavelength limit ($q=0$),
all with $G_0=0.8$~\cite{RGnm}, and in the absence of mean-field effects
($G_0=0$).\label{rms_energy_change}}
\end{figure}

{\it Energy transfer.} Figure~\ref{rms_energy_change} shows the impact
of $C_\sigma$ on the root-mean-square (rms) energy transfer $\Delta E$
in scattering from nucleons. In the absence of Pauli blocking of final
neutrino states, $\Delta E$ is given by
\be
(\Delta E)^2 = \frac{\int d{\bf p}'_\nu \, (E_\nu-E'_\nu)^2 \, 
\Gamma(E_\nu-E'_\nu,p_\nu-p'_\nu)}{\int d{\bf p}'_\nu \, 
\Gamma(E_\nu-E'_\nu,p_\nu-p'_\nu)} \,.
\ee
For the N$^3$LO results, the comparison with the long-wavelength approximation
($q=0$) shows that the energy transfer due to nucleon recoil is small.
Collision processes are the major contributor to the energy transfer.
This is also clear from the decreased $\Delta E$ with the lower $C_\sigma$
value. In addition, the comparison with the $G_0=0$ results demonstrates
the increased energy transfer due to repulsive mean-field effects in the
spin channel.

\begin{figure}[t]
\begin{center}
\includegraphics[scale=0.44,clip=]{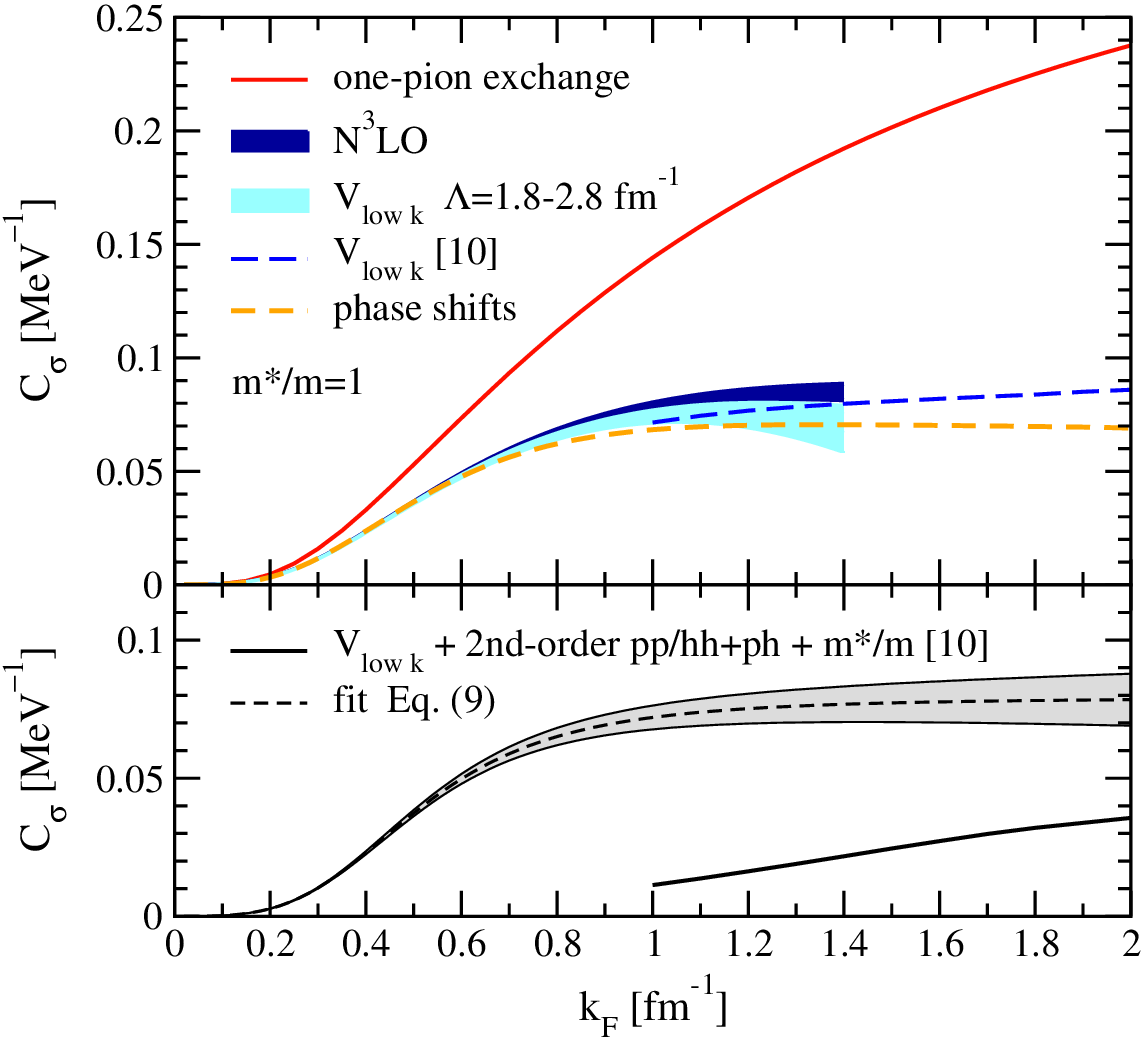}
\end{center}
\caption{(Color online) Combined results for the spin relaxation
rate given by $C_\sigma$: bands for different chiral N$^3$LO 
potentials~\cite{EGM,EM} and chiral $\vlowk$ interactions~\cite{Vlowk}
and the result based on phase shifts. We also include the rate
of Ref.~\cite{LPS} for $\vlowk$ with a density-dependent cutoff
$\Lambda = \sqrt{2} \, k_{\rm F}$. All results are for $m^*/m = 1$
(also for Ref.~\cite{LPS}). Lower panel: The band represents the 
$C_\sigma$ range in the upper panel based on nuclear 
interactions and NN phase shifts. The solid line includes 
second-order many-body and $m^*$ contributions, with $m^*/m 
\approx 1.12 - 0.17 \, (\kf/\fmi)$ over this range~\cite{LPS}.
\label{combined}}
\end{figure}

{\it Comparison of interactions.} We have also carried out calculations 
for the Entem and Machleidt (EM) N$^3$LO potentials with $\Lambda = 500$
and $600 \mev$~\cite{EM}. The EM $500 \mev$ results overlap with the 
EGM N$^3$LO band of Fig.~\ref{orders}. The EM $600 \mev$ rate is larger
due to the failure of the Born approximation in this case. We have 
also used the renormalization group (RG)~\cite{Vlowk} to evolve all
N$^3$LO~\cite{EGM,EM} potentials to low-momentum interactions
$\vlowk$ with $\Lambda=1.8$--$2.8 \fmi$ ($360$--$560 \mev$) to extend
the estimate of the theoretical uncertainty. The RG preserves the 
long-range pion exchanges and includes subleading contact interactions,
so that NN scattering data are reproduced. The resulting $C_\sigma$ band
is shown in the upper panel of Fig.~\ref{combined}. This includes the
rate based on the evolved EM $600 \mev$ potential.

In the low-density limit, two-nucleon collisions dominate and NN phase
shifts provide a model-independent result for the spin relaxation
rate. In the near-degenerate case, the strong potential $V$ in
Eq.~(\ref{pw}) is then replaced by the free-space $K$ matrix, since
Pauli blocking removes the imaginary parts from the loop integrals in
the $T$-matrix equation. We use the standard expression of the $K$
matrix in terms of empirical phase shifts and mixing angles~\cite{BJ}
from the Nijmegen Partial Wave analysis (nn-online.org). In general,
it is however unclear whether an expansion in terms of the free-space
$K$ matrix (or the $T$ matrix) is reliable for $\kf \gtrsim 0.1 \fmi$
due to large scattering lengths and Pauli blocking for near-degenerate
conditions, unless the relevant parts of $V$ are
perturbative. Neutrino-pair bremsstrahlung neglecting mean-field and
LPM effects has been calculated from the $T$ matrix in
Ref.~\cite{Hanhart}. For all densities in Fig.~\ref{combined}, we find
the striking result that the chiral EFT and $\vlowk$ rates obtained in
Born approximation are close to those from phase shifts (with the $K$
matrix). This demonstrates that, for these lower cutoffs, the
noncentral part of the neutron-neutron amplitude is perturbative in
the particle-particle channel.

The upper panel of Fig.~\ref{combined} combines our results
based on nuclear interactions and NN phase shifts, and also
includes rates from $\vlowk$ interactions~\cite{LPS} that
extend to higher densities. At subnuclear densities $\rho < 
10^{14} \gcmiq$ ($\kf < 1.2 \fmi$), the spin response is well
constrained and all results lie within a band, with a 
significantly reduced $C_\sigma$ compared to OPE.
For $\kf \gtrsim 1.5 \fmi$, calculations of the equation of
state show that low-momentum 3N interactions become 
important~\cite{neutmatt} and should be included. In the
lower panel of Fig.~\ref{combined}, we show a simple
fit representing our results,
\be
\frac{C_\sigma}{{\rm MeV}} = \frac{0.86 \, (\kf/{\rm fm}^{-1})^{3.6}}{1 
+ 10.9 \, (\kf/{\rm fm}^{-1})^{3.6}} \,,
\label{fit}
\ee
that, together with the spin response function given by
Eqs.~(\ref{structspin}) and~(\ref{Xsigma}), can be used in astrophysical 
simulations. 

{\it Many-body effects.} Our results above do not include the
effects of the nuclear medium on collisions. To give a sense
of how these affect $C_\sigma$, we show in the lower panel of
Fig.~\ref{combined} rates that include second-order many-body
contributions (particle-particle/hole-hole and particle-hole)
and self-energy effects through the effective mass $m^*$ (according
to Eq.~(\ref{pw}), $C_\sigma \sim m^{* \, 3}$)~\cite{LPS}.
Both particle-hole and $m^*$ effects
reduce the spin relaxation rate. The former is driven by 
second-order particle-hole mixing of tensor with strong
central interactions~\cite{noncentral}. This demonstrates
the need to study in greater detail the influence of
many-body effects on collisions (see also Ref.~\cite{Dalen}).

In summary, we have presented the first calculations of neutrino
processes in supernovae based on chiral EFT. Our N$^3$LO results
over the important density range $\rho \lesssim 10^{14} \gcmiq$
represent a significant advance beyond the OPE rates currently
used in simulations. Shorter-range noncentral forces reduce the 
rates significantly. For densities $\rho < 10^{14} \gcmiq$, the
spin response is well constrained by nuclear interactions and NN
scattering data. Future work will include neutron-proton mixtures
and charged currents, to systematically improve the neutrino 
physics input for astrophysics.

\begin{acknowledgments}
We thank S.\ Bogner, M.\ Liebend{\"o}rfer, and A.\ Nogga 
for useful discussions, and the Niels Bohr International Academy
and NORDITA for their hospitality. This work was supported in 
part by NSERC and the NRC Canada.
\end{acknowledgments}

\end{document}